\documentclass[aps,prl,twocolumn,showpacs]{revtex4}
\usepackage{amsmath}
\usepackage{epsfig}

\def\C{\hbox{$\mit I$\kern-.7em$\mit C$}}
\def\R{\hbox{$\mit I$\kern-.7em$\mit R$}}
\newcommand{\id}{\mbox{$1 \hspace{-1.0mm}  {\bf l}$}}
\renewcommand{\Re}{{\mathrm{Re}}}
\renewcommand{\Im}{{\mathrm{Im}}}
\def\Box{\rule{1ex}{1ex}}
\def\Proof{\textsc{Proof: }}
\newtheorem{theorem}{Theorem}
\newtheorem{proposition}{Proposition}
\newtheorem{lemma}{Lemma}
\newtheorem{corollary}{Corollary}
\begin{document}

\title{Separability Criterion for all bipartite Gaussian States}

\author{G. Giedke$^{(1)}$, B. Kraus$^{(1)}$, M. Lewenstein$^{(2)}$, and J. I. Cirac$^{(1)}$}

\affiliation{(1) Institut f\"ur Theoretische Physik, Universit\"at Innsbruck,
A-6020 Innsbruck, Austria\\
(2) Institut f\"ur Theoretische Physik, Universit\"at
Hannover, 30163 Hannover, Germany}

\date{\today}
  
\begin{abstract}
We provide a necessary and sufficient condition for separability
of Gaussian states of bipartite systems of arbitrarily many
modes. The condition provides an operational criterion since it
can be checked by simple computation. Moreover, it allows us to
find a pure product--state decomposition of any given separable
Gaussian state. Our criterion is independent of the one based on
partial transposition, and is strictly stronger.
\end{abstract}


\maketitle


Entanglement is the basic ingredient in the philosophical
implications of Quantum Theory. It also plays a crucial role in
some fundamental issues of this theory, such as decoherence or
the measurement process. Furthermore, it is the basis of most
applications in the field of Quantum Information. However,
despite of its importance, the entanglement properties of
systems are far from being understood. In particular, we do not
even know how to solve the following question \cite{WernerSep}:
given two systems A and B in a state described by a density
operator $\rho$, are those systems entangled? This question
constitutes the so--called separability problem, and it
represents one of the most important theoretical challenges of
the emerging theory of quantum information.

During the last few years a significant amount of work in the
field of quantum information has been devoted to the
separability problem \cite{SepGen}. For the moment, the basic
tool to study this problem is a {\em linear} map called partial
transposition operation. Introduced in this context by Peres
\cite{QBSepCrit1}, it provides us with a necessary condition for
a density operator to be separable (equivalently, not
entangled). This condition turns out to be sufficient as well
for two particular cases: (a) A and B are two qubits or one
qubit and one qutrit \cite{QBSepCrit2}; (b) A and B are two
modes (continuous variable systems) in a Gaussian state
\cite{Duan99}. Thus, in these two cases the separability problem
can be fully solved. However, for higher dimensional systems as
well as in the case in which A and B consist of several modes in
a joint Gaussian state, partial transposition alone does not
provide a general criterion for separability. In both cases,
examples of states which despite of being entangled satisfy the
partial transposition criterion have been provided
\cite{BE,We01}.

In this Letter we solve the problem of separability for Gaussian
states of an arbitrary number of modes per site. Our method does
not rely in any sense on the concept of partial transposition,
and therefore is diametrically different from the ones that have
been introduced so far to study the separability problem
\cite{SepGen}. It is based on a {\em nonlinear} map
$f\!:\gamma_N\to\gamma_{N+1}$ between matrices $\gamma_N$ which
reveals whether a state $\rho$ is entangled state or not.
Furthermore, once $\rho$ is shown to be separable, our method
allows to find an explicit decomposition of $\rho$ as a convex
combination of product states.

Let us start by fixing the notation and recalling some
properties of correlation matrices (CMs). A Gaussian state of
$n$ modes is completely characterized by a matrix $\gamma \in
M_{2n,2n}$ (the set of $2n\times 2n$ matrices), called
correlation matrix \cite{GenCV}, whose elements are directly
measurable quantities. A matrix $\gamma\in M_{2n,2n}$ is a CM if
it is real, symmetric, and $\gamma - iJ_n\ge 0$. Here
\cite{note_dirsum}
\begin{equation}
J_n\equiv \oplus_{k=1}^n J_1,\quad J_1 \equiv
\left(\begin{array}{cc} 0 & -1 \\ 1 & 0 \end{array}\right).
\end{equation}

In the following we will consider two systems A and B, composed
of $n$ and $m$ modes, respectively, in a Gaussian state. The
corresponding CM will be written as
\begin{equation}
\label{Eq_gamma}
\gamma_0 = \left(\begin{array}{cc} A_0 & C_0 \\
C_0^T & B_0 \end{array}\right) \ge iJ_{n,m}
\end{equation}
where $A_0\in M_{2n,2n}$ and $B_0\in M_{2m,2m}$ are CM
themselves, $C_0 \in M_{2n,2m}$ and $J_{n,m}\equiv J_n\oplus
J_m$. In order to simplify the notation, when it is clear from
the context we will not write the subscripts to the matrices $J$
and we will not specify the dimensions of the matrices involved
in our derivations. In \cite{We01} it was shown that a CM of the
form (\ref{Eq_gamma}) is separable (i.e., it corresponds to a
separable state) iff there exist two CMs, $\gamma_{A,B}$, such
that
\begin{equation}
\label{SepCond}
\gamma_0 \ge \gamma_A\oplus \gamma_B.
\end{equation}
This condition, even though it can be very useful to show that
some particular states are entangled \cite{We01,trient}, cannot
be directly used in practice to determine whether an arbitrary
state is entangled or not, since there is no way of determining
$\gamma_{A,B}$ in general. If one can determine them, however,
then one can automatically construct an explicit decomposition
of the corresponding density operator as a convex combination of
product states \cite{We01}.

Below we will present a criterion which allows us to determine
whether a given CM, $\gamma_0$, is separable or not, and which
allows us to determine a product--state decomposition if this is
the case. To this aim, we define a sequence of matrices
$\{\gamma_N\}_{N=0}^\infty$ of the form (\ref{Eq_gamma}). The
matrix $\gamma_{N+1}$ can be determined through the discrete map
defined as follows: (i) if $\gamma_N$ is not a CM then
$\gamma_{N+1}=0$; (ii) if $\gamma_N$ is a CM then
\begin{subequations}
\label{Eq_map}
\begin{eqnarray}
A_{N+1}\equiv B_{N+1} &\equiv& A_N - \Re (X_N), \\
 C_{N+1} &\equiv& -\Im (X_N),
\end{eqnarray}
\end{subequations}
where $X_N \equiv C_N (B_N-iJ)^{-1} C_N^T$ \cite{note_pinv}.
Note that for $N\ge 1$ we have that $A_N=A_N^T=B_N$ and
$C_N=-C_N^T$ are real matrices. The importance of this sequence
is that, as we will show below, $\gamma_0$ is separable iff
$\gamma_N$ is a valid separable CM. In particular, for some
finite number of iterations $\gamma_N$ will acquire a form in
which it is simple to check that it is separable. Furthermore,
starting from that CM we will be able to construct the CMs
$\gamma_{A,B}$ of Eq.\ (\ref{SepCond}) for the original
$\gamma_0$. Now we will present several propositions from which
the above results will follow. Two lemmas are presented in an
appendix.

First we show that if $\gamma_N$ is separable, so is
$\gamma_{N+1}$. Moreover, the CMs $\gamma_{A,B}$ associated to
$\gamma_N$ [cf. Eq.\ (\ref{SepCond})] allow us to construct the
corresponding CMs for $\gamma_{N+1}$.

\begin{proposition}
\label{proposition2}
If for some CMs, $\gamma_{A,B}$, we have $\gamma_N\ge
\gamma_A\oplus\gamma_B$ then $\gamma_{N+1}\ge \gamma_A\oplus\gamma_A$.
\end{proposition}

\Proof\
We use the equivalence (i)--(iii) of Lemma \ref{Lemma1} to
obtain that $B_N-C_N^T(A_N-\gamma_A)^{-1}C_N\ge \gamma_B\ge iJ$,
where the last inequality follows from the fact that $\gamma_B$
is a CM. Using the equivalence (ii)--(iii) of Lemma \ref{Lemma1}
we obtain $\gamma_A\le A_N-C_N(B_N-iJ)^{-1}C_N^T=A_{N+1} +i
C_{N+1}$, where we have also used the map (\ref{Eq_map}).
According to Lemma \ref{Lemma2}, this immediately proves the
proposition.
\hfill\Box

Now, we show that the converse of Prop.\ \ref{proposition2} is
true. That is, if $\gamma_{N+1}$ is separable, so is $\gamma_N$.
Apart from that, the following proposition exhibits how to
construct the matrices $\gamma_{A,B}$ [cf. Eq.\ (\ref{SepCond})]
related to $\gamma_N$ starting from the ones corresponding to
$\gamma_{N+1}$.

\begin{proposition}
\label{proposition3}
If for some CM, $\gamma_{A}$, we have that $\gamma_{N+1}\ge
\gamma_A\oplus\gamma_A$ then $\gamma_{N}\ge \gamma_A\oplus\gamma_B$,
where
\begin{equation}
\label{Eq_Result2b}
\gamma_B \equiv B_N - C_N (A_N-\gamma_A)^{-1} C_N^T,
\end{equation}
is a CM.
\end{proposition}

\Proof\
We use Lemma \ref{Lemma2} and the map (\ref{Eq_map}) to
transform the inequality $\gamma_{N+1}\ge
\gamma_A\oplus\gamma_A$ into $A_N - C_N^T (B_N-iJ)^{-1} C_N
\ge \gamma_A$. According to the equivalence (ii)--(iii)
of Lemma \ref{Lemma1} this implies that $\gamma_B\ge iJ$. Since
it is clear from its definition (\ref{Eq_Result2b}), $\gamma_B$
is also real and symmetric, it is a CM. On the other hand, using
the equivalence (i)--(iii) of Lemma \ref{Lemma1} we immediately
obtain that $\gamma_N\ge \gamma_A\oplus\gamma_B$.\hfill\Box

Using the fact that for $N\ge 1$, $A_N=B_N$ and the symmetry of
the corresponding matrix $\gamma_N$ we have
\begin{corollary}
\label{corollary1}
Under the conditions of Prop.\ \ref{proposition3}, if $N\ge 1$
we have that $\gamma_{N}\ge
\tilde\gamma_A\oplus\tilde\gamma_A$, where $\tilde\gamma_A\equiv
(\gamma_A+\gamma_B)/2\ge iJ$ is a CM.
\end{corollary}

The above propositions imply that $\gamma_0$ is separable iff
$\gamma_N$ is separable for all $N>0$. Thus, if we find some
$\gamma_N$ fulfilling (\ref{SepCond}) then $\gamma_0$ is
separable. Thus, we can establish now the main result of this
work.

\begin{theorem}\label{theorem1} (Separability criterion)

\begin{description}

\item [(1)] If for some $N\ge 1$ we have $A_N\not \ge iJ$ then
$\gamma_0$ is not separable.

\item [(2)] If for some $N\ge 1$ we have
\begin{equation}
\label{Eq_condLN}
L_N\equiv A_N- ||C_N||_{\rm op} \id \ge iJ
\end{equation}
then $\gamma_0$ is separable \cite{note_norms}.
\end{description}
\end{theorem}

\Proof\
(1) It follows directly from Prop.\ \ref{proposition2}; (2) We
will show that $\gamma_N\ge L_N\oplus L_N$, so that according to
Prop.\ \ref{proposition3} $\gamma_0$ is separable. We have
\begin{equation}
\gamma_N=L_N\oplus L_N +
\left(\begin{array}{cc} ||C_N||_{\rm op} \id & C_N\\
C_N^T & ||C_N||_{\rm op} \id \end{array}\right),
\end{equation}
so that we just have to prove that the last matrix is positive.
But using Lemma \ref{Lemma1} this is equivalent to $||C_N||_{\rm
op}^2\id\ge C_N^TC_N$, which is always the case.\hfill\Box

This theorem tells us how to proceed in order to determine if a
CM is separable or not. We just have to iterate the map
(\ref{Eq_map}) until we find that either $A_N$ is no longer a CM
or $L_N$ is a CM. In the first case, we have that $\gamma_0$ is
not separable, whereas in the second one it is separable. If we
wish to find a decomposition of the corresponding density
operator as a convex set of product vectors we simply use the
construction given in Corollary \ref{corollary1} until $N=1$ and
then the one of Prop.\ \ref{proposition3}. This will give us the
CMs, $\gamma_{A,B}$, such that $\gamma_0\ge
\gamma_A\oplus\gamma_B$, from which the decomposition can be
easily found \cite{We01}.

\begin{figure}[tbp]
\epsfxsize=8.5cm
\epsfysize=3.8cm
\begin{center}
\epsffile{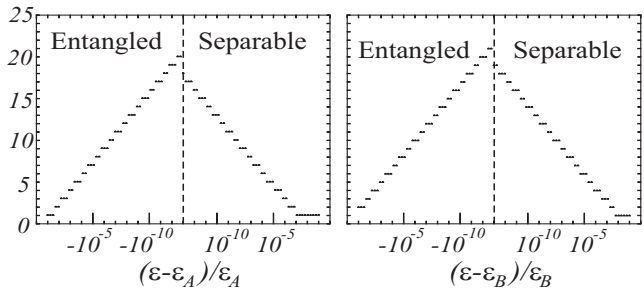}
\end{center}
\caption{\label{fig1}
Number of steps as a function $\epsilon$ for CMs of the form
$\gamma_{a,b}(\epsilon)=\gamma_{a,b}+\epsilon\id$ where: (a)
$\gamma_a$ taken from Eq.\ (1) in Ref.\ \cite{GHZ} with $r=1/4$,
and $\epsilon_a=0.305774915510(1)$; (b) $\gamma_b$ taken from
Eq.\ (9) in Ref.\ \cite{We01} and
$\epsilon_b=0.0978667902228(4)$.}
\end{figure}

In order to check how fast our method converges we have taken
families of CMs and applied to them our criterion. We find that
typically with less than 5 iterations we are able to decide
whether a given CM is entangled or not. The most demanding
states for the criterion are those which lie very close to the
border of the set of separable states (see Corollary
\ref{corollary2} below). We challenged the criterion by applying
it to states close to the border of the set of separable states
and still the convergence was very fast (always below 30 steps).
Figure \ref{fig1} illustrates this behavior. We have taken
$n=m=2$ modes, an entangled CM $\gamma_{a}$ of the GHZ form
\cite{GHZ} (Fig. \ref{fig1}a) and an entangled CM $\gamma_b$
with positive partial transposition \cite{We01} (Fig.
\ref{fig1}b). We produced two families of CMs as
$\gamma_{a,b}(\epsilon)=\gamma_{a,b}+\epsilon\id$. We have
determined $\epsilon_{a,b}$ such that the CMs become separable.
In the figure we see that in both cases, as we approach
exponentially fast $\epsilon_{a,b}$, the number of steps only
increases linearly. We have also added, instead of $\id$ other
positive projectors with all possible ranks and found the same
behavior. By taking other initial CMs we also find the same
results.

Even though we have tested numerically the rapid convergence of
our method, we still have to prove that, except for a zero
measure set, it can decide whether a CM is entangled or not
after a finite number of steps \cite{note_ppt}. We start out by
considering the set of separable states. Since for them
$\gamma_0\ge \gamma_A\oplus\gamma_B$ with $\gamma_{A,B}\ge iJ$,
if we just consider those with $\gamma_A > iJ$, we will leave
out a zero measure set. In this case we can show that after a
finite number of steps these separable states will be detected
by using our procedure.

\begin{proposition}
\label{proposition4}
If $\gamma_0\ge \gamma_A\oplus\gamma_B$ with $\gamma_A\ge iJ+\epsilon
\id$, then there exists some
\begin{equation}
N<N_0\equiv \frac{1}{\epsilon}(||A_0||_{\rm tr}-2n)+1,
\end{equation}
for which condition (\ref{Eq_condLN}) is fulfilled.
\end{proposition}

\Proof\
Using Prop.\ \ref{proposition2} we have that for all $N$,
\begin{equation}
\label{Eq_prop4}
A_N-iJ\ge \epsilon\id.
\end{equation}
Thus, $0\le \Re(X_N)=A_N-A_{N+1}$. Since all the matrices in
this expression are positive, taking the trace norm we have
$||A_N||_{\rm tr}-||A_{N+1}||_{\rm tr} = ||\Re(X_N)||_{\rm tr}$.
Adding both sides of this equation from $N=0$ to $N_0$, taking
into account that $||\ldots||_{\rm tr}\ge ||\ldots||_{\rm op}$,
and $||\Re(X_N)||_{\rm op}
\ge ||C_{N+1}||_{\rm op}$ [since $\Re(X_N)\ge \pm i \Im(X_N)$], we have
\begin{equation}
\sum_{N=0}^{N_0-1} ||C_{N+1}||_{\rm op} \le ||A_0||_{\rm tr}-||A_{N_0}||_{\rm tr}
\le ||A_0||_{\rm tr}-2n,
\end{equation}
where the last inequality is a consequence of the fact that
$A_N\ge iJ$ for all $N$. Thus, among $\{C_{N}\}_{N=1}^{N_0}$
there must be at least one for which $||C_N||_{\rm op}\le
\epsilon$. Thus, $A_N-||C_N||_{\rm op}\id\ge A_N-\epsilon\id\ge
0$ where for the last inequality we have used Eq.\
(\ref{Eq_prop4}), and therefore, for that particular value of
$N$, condition (\ref{Eq_condLN}) must be fulfilled.\hfill\Box

It is worth stressing that from the proof of Prop.\
\ref{proposition4} it follows directly that if $\gamma_0$ is
separable, then the sequence $\gamma_N$ converges to a fixed
point $\gamma_{\infty}=A_\infty\oplus B_\infty$, where
$A_\infty=B_\infty\ge iJ$ are CMs. On the other hand, for the
sake of completeness, we will now show that if $\gamma_0$ is not
separable, then we can always detect it in a finite number of
steps. We will use the fact that the CMs of inseparable Gaussian
states form an open set, a fact that can be directly inferred
from condition (\ref{SepCond}). This means that if $\gamma_0$ is
inseparable, there always exist some $\epsilon_0>0$ such that if
$\epsilon<\epsilon_0$ then $\gamma_0 + \epsilon \id$ is still
inseparable and therefore condition (\ref{Eq_condLN}) is never
fulfilled. However, if $\gamma_0$ was separable, then, according
to Prop.\ \ref{proposition4}, $\gamma_0 + \epsilon \id$ should
fulfill that condition before reaching $N=N_0$. This can be
summarized as follows.

\begin{corollary}
\label{corollary2}
If $\gamma$ is inseparable then there exists some $\epsilon>0$
such that starting out from $\gamma_0=\gamma+\epsilon\id$,
condition (\ref{Eq_condLN}) is not fulfilled for any $N\le
N_0\equiv (||A_0||_{\rm tr}-2n)/\epsilon$.
\end{corollary}

Together, Prop.\ \ref{proposition4} and Corollary
\ref{corollary2} indicate that whether $\gamma_0$ is separable
or not, and except for a set of zero measure, we will be able to
detect it in a finite number of steps. However, as mentioned
above, according to our numerical calculations we see that the
process always converges very fast and in practice one can
directly use the method sketched after Theorem 1.

In conclusion, we have obtained a necessary and sufficient
condition for Gaussian states to be separable. The condition
provides an operational criterion in that it can be easily
checked by direct computation. It is also worth mentioning that
our criterion can be used to study the separability properties
with respect to bipartite splitting of multipartite systems in
Gaussian states \cite{Dur}. Our criterion is based on a
non--linear map, and is more powerful than partial
transposition. This fact indicates that in other situations,
like the one in which the systems A and B are $n$ and $m$--level
systems with $n\times m>6$, there might exist a more powerful
criterion than partial transposition to determine whether states
are separable or not. This problem still remains open. However,
the results presented here represent a significant step in
understanding the separability problem, which is one of the most
challenging problems in the field of quantum information.

G.G. acknowledges financial support by the
Friedrich-Naumann-Stiftung. This work was supported by the
Austrian Science Foundation (SFB ``Control and Measurement of
Coherent Quantum Systems'', Project 11), the EU (EQUIP, contract
IST-1999-11053), the ESF, the Institute for Quantum Information
GmbH in Innsbruck, and the DFG (SFB 407 and Schwerpunkt
"Quanteninformationsverarbeitung").

\noindent
\appendix
\subsection*{Appendix}

In this Appendix we present the lemmas which are needed in order
to prove Props.\ \ref{proposition2} and \ref{proposition3}.

Let us consider three real matrices $0\le A=A^T\in M_{n,n}$,
$0\le B=B^T\in M_{m,m}$, and $C\in M_{n,m}$, and
\begin{equation}
M = \left(\begin{array}{cc} A & C \\ C^T & B\end{array}\right)
=M^T \in M_{n+m,n+m}.
\end{equation}

\begin{lemma}
\label{Lemma1}
The following statements are equivalent:

\begin{description}

\item [(i)] $M\ge 0$.

\item [(ii)] ${\rm ker}(B)\subseteq {\rm ker}(C)$ and $A-CB^{-1}C^T\ge
0$.

\item [(iii)] ${\rm ker}(A)\subseteq {\rm ker}(C^T)$ and
$B-C^TA^{-1}C\ge 0$ \cite{note_pinv}.
\end{description}

\end{lemma}

\Proof\
We will just prove the first equivalence since the other one is
analogous. We use that $M\ge 0$ iff for any two real vectors
$a\in \R^n$ and $b\in\R^m$
\begin{equation}
\label{Lemma1a}
a^TAa+b^TBb+a^TCb+b^TC^Ta\ge 0.
\end{equation}
On the other hand, $A-CB^{-1}C^T\ge 0$ iff for any $a\in \R^n$
we have
\begin{equation}
\label{Lemma1b}
a^TAa-a^TCB^{-1}C^Ta\ge 0.
\end{equation}
(i)$\Rightarrow$(ii): We assume (\ref{Lemma1a}). First, ${\rm
ker}(B)\subseteq {\rm ker}(C)$ since otherwise we could always
choose a $b\in{\rm ker}(B)$ so that $-a^TCb>a^TAa$. Second, if
we choose $b=-B^{-1}C^Ta$ then we obtain (\ref{Lemma1b}).
(ii)$\Rightarrow$(i): We now assume (\ref{Lemma1b}). Then,
$A=CB^{-1}C^T+P$, where $P\ge 0$. Defining $\tilde a\equiv
B^{-1}C^Ta$, we have that $C^Ta=B\tilde a$ (since ${\rm
ker}(B)\subseteq {\rm ker}(C)$), and thus the lhs of
(\ref{Lemma1a}) can be expressed as $a^TPa+(\tilde
a+b)^TB(\tilde a+b)$, which is positive.\hfill\Box

In the derivations of Props.\ \ref{proposition2} and
\ref{proposition3} we have not included explicitly the
conditions imposed by the present lemma on the kernels of $B$
and $C$. However, one can easily verify that all the problems
that may arise from these kernels are eliminated by using
pseudoinverses instead of inverses of matrices \cite{note_pinv}.

Let us consider two real matrices $A=A^T\in M_{n,n}$ and
$C=-C^T\in M_{n,n}$, and
\begin{equation}
M= \left(\begin{array}{cc} A & C\\ C^T & A\end{array}\right)
=M^T
\in M_{2n,2n}.
\end{equation}

\begin{lemma}
\label{Lemma2}
$M\ge 0$ iff $A+iC\ge 0$.
\end{lemma}

\Proof\
This follows from the observation that $M$ is real, and that for
any pair of real vectors $a,b\in \R^N$ we have $(a-ib)^\dagger
(A+iC)(a-ib)=(a\oplus b)^T M (a\oplus b)$.\hfill\Box



\begin{thebibliography}{99}

\bibitem{WernerSep}
R.F. Werner,
Phys. Rev. A \textbf{40}, 4277 (1989).

\bibitem{SepGen}
For a review of the problem and its progress see, for example,
M. Lewenstein, D. Bruss, J.I. Cirac, B. Kraus, M. Ku\'s, J.
Samsonowicz, A. Sanpera, and R. Tarrach, in J. Mod. Opt.
\textbf{47}, 2481 (2000).

\bibitem{QBSepCrit1}
A. Peres, Phys. Rev. Lett. {\bf 77}, 1413 (1996).

\bibitem{QBSepCrit2}
M., P., and R. Horodecki, Phys. Lett. A \textbf{223}, 1 (1996).

\bibitem{Duan99} L.-M. Duan, G. Giedke, J.I. Cirac, and P. Zoller,
Phys. Rev. Lett. \textbf{84}, 2722 (2000);
R. Simon,
Phys. Rev. Lett. \textbf{84}, 2726 (2000).

\bibitem{BE}
P. Horodecki, Phys. Lett. A \textbf{232}, 333 (1997); C.H.
Bennett, D.P. DiVincenzo, T. Mor, P.W. Shor, J.A. Smolin, and
B.M. Terhal, Phys. Rev. Lett. \textbf{82}, 5385 (1999).

\bibitem{We01}
R.F. Werner and M.M. Wolf, to appear in Phys. Rev. Lett.
\textbf{86}; quant-ph/0009118 (2000).

\bibitem{GenCV}
If $X_k,P_k$ are position- and momentum--like operators in each
mode fulfilling canonical commutation relation $[X_k,P_k]=i$, we
define
\[
\gamma_{\alpha,\beta}\equiv 2 \Re[\langle (R_\alpha-d_\alpha)
(R_\beta-d_\beta)\rangle],
\]
where $d_\alpha=\langle R_\alpha\rangle$ and $R_{2k-1}=X_k$ and
$R_{2k}=P_k$ ($k=1,2,\ldots,n$).

\bibitem{note_dirsum}
For convenience we use direct sum notation for matrices and vectors.
That is, if $A\in M_{n,n}$ and $B\in M_{m,m}$, $A\oplus B\in M_{n+m,
n+m}$ is a block diagonal matrix of blocks $A$ and $B$. Similarly,
if $f_1\in \R^n$ and $f_2\in \R^m$ are two vectors, then $f_1\oplus f_2
\in \R^{n+m}$ is a vector whose first $n$ components are given by the
entries of $f_1$ and the last $m$ by those of $f_2$.

\bibitem{trient}
G. Giedke, B. Kraus, J.I. Cirac, and M. Lewenstein,
quant-ph/01030137.

\bibitem{note_pinv}
Throughout this work we will denote by $B^{-1}$ the pseudoinverse of $B$, that is,
$BB^{-1}=B^{-1}B$ is the projector on the range of $B$. If $B$ is invertible,
$B^{-1}$ coincides with the inverse of $B$.

\bibitem{note_norms}
$||A||_{\rm tr}\equiv {\rm tr}(A^\dagger A)^{1/2}$ denotes the
trace norm of $A$. The operator norm of $A$, $||A||_{\rm op}$ is
the maximum eigenvalue of $(A^\dagger A)^{1/2}$.

\bibitem{GHZ}
P. van Loock and S. L. Braunstein, Phys. Rev. A \textbf{63},
022106 (2001).

\bibitem{note_ppt}
Note that the fact that there exists a zero measure set which is
not possible to characterize in a finite number of steps is not
particular for our method, but a simple consequence of finite
precision. For example, if we have a density operator $\rho$ for
two qubits such that the partial transpose has one negative
eigenvalue $-\epsilon$, it will be increasingly difficult to
check whether $\rho^T\ge 0$ as $\epsilon\to 0$.

\bibitem{Dur}
W. D\"ur, J.I. Cirac, and R. Tarrach, Phys. Rev. Lett.
\textbf{83}, 3562 (1999); W. D\"ur and J.I. Cirac, Phys. Rev. A
\textbf{61}, 042314 (2000).



\end{thebibliography}
\end{document}